\documentclass[10pt,a4paper]{article} % Paper type (a4paper, usletter or legal) and font size (10, 11 or 12)

\usepackage{charter} % Charter font for main content

\usepackage{graphicx} % Required for including images
\usepackage{amssymb,amsmath} % Math packages
\usepackage{multicol} % Required for the three-column layout of the document
\usepackage{url} % Clickable links
\usepackage{enumitem} % Reduces the amount of space within and between lists with [noitemsep,nolistsep]
\usepackage{marvosym} % Required for the use of symbols
\usepackage{wrapfig} % Allows wrapping text around figures
\usepackage[T1]{fontenc} % Use 8-bit encoding that has 256 glyphs
\usepackage{datetime} % Required for defining a custom date style
\newdateformat{mydate}{\monthname[\THEMONTH] \THEYEAR} % Set a custom date format
\usepackage[hidelinks]{hyperref}
\usepackage{fancyhdr} % Required to define custom headers/footers
\usepackage{xcolor}
\newcommand{\NewsletterName}[1]{ % Newsletter title
\begin{center}
\Huge \usefont{T1}{fvs}{b}{n} % Use the Bera Sans Bold font
#1
\end{center}	
\par \normalsize \normalfont}

\newcommand{\NewsItem}[1]{ % News item title
\usefont{T1}{fvs}{n}{n} % Use the Bera Sans Normal font
\vspace{24pt}\large #1\vspace{3pt} % Print the title with space around it in a larger font size
\par \normalsize \normalfont}

\newcommand{\NewsAuthor}[1]{ % Author name under the item title
\hfill by \textsc{#1} \vspace{20pt} % Right-aligned author name in small caps with space after it
\par \normalfont}		

\begin{document}

%\JournalIssue{4} % Issue number

\NewsletterName{Towards a fast and robust deep hedging approach} % Newsletter title
\NewsAuthor{Fabienne Schmid\footnote{ RIVACON GmbH, f.schmid@rivacon.com}, Daniel Oeltz\footnote{Fraunhofer SCAI, daniel.oeltz@scai.fraunhofer.de}} 

%\noindent\HorRule{3pt} \\[-0.75\baselineskip] % Thick horizontal rule
%\HorRule{1pt} % Thin horizontal rule

%----------------------------------------------------------------------------------------
%	MAIN NEWS ITEM
%----------------------------------------------------------------------------------------

%\vspace{0.5cm}
%\SepRule
\vspace{-1cm}

\begin{center}
\begin{minipage}[h]{\linewidth}
\NewsItem{\textbf{Summary}}% Main next item title
\vspace{0.3cm} % Some extra whitespace since there is no author as for the news in the body of the newsletter
\noindent
\textit{We present a robust Deep Hedging framework for the pricing and hedging of option portfolios that significantly improves training efficiency and model robustness. In particular, we propose a neural model for training model embeddings which utilizes the paths of several advanced equity option models with stochastic volatility in order to learn the relationships that exist between hedging strategies.  A key advantage of the proposed method is its ability to rapidly and reliably adapt to new market regimes through the recalibration of a low-dimensional embedding vector, rather than retraining the entire network. Moreover, we examine the observed Profit and Loss distributions on the parameter space of the models used to learn the embeddings. The results show that the proposed framework works well with data generated by complex models and can serve as a construction basis for an efficient and robust simulation tool for the systematic development of an entirely model-independent hedging strategy.
}
\end{minipage}
\end{center}

\vspace{0.2cm}
%\SepRule % Small horizontal rule after the main news item
%\vspace{0.5cm}

%\setlength{\columnsep}{16pt} % Uncomment to manually change the white space between columns
\begin{multicols}{2} % Begin the three-column layout

%----------------------------------------------------------------------------------------
%	OTHER NEWS
%----------------------------------------------------------------------------------------
\vspace{-0.3cm}
\section{Introduction}
%\NewsAuthor{Author Name}
\vspace{-0.3cm}
Recent progress in deep learning has achieved remarkable results in many areas, such as computer visions, image classification or natural language processing. Especially, with the increase in the availability of observational and numerical datasets and advances in capabilities of computational power, complex and time-intensive algorithms are now feasible for various practical uses. In particular, deep neural network models excel at processing vast amounts of data as well as identifying complex relationships embedded in the data, and are perfectly suited for addressing optimization problems. 
\\
Although, deep learning was developed in the field of computer science, its application has gained increasing attention within the financial sector, too, leading to a growing number of deep learning models that can provide real-time working solutions, such as stock market forecasting or credit risk assessment (cf. \cite{ozbayoglu2020deep}). 
A special case of the practical application of deep learning algorithms in finance and banking lies in hedging derivatives (so-called \textit{Deep Hedging}), where a neural network is used to derive hedging strategies for a portfolio of derivatives by minimizing a chosen risk measure. Although, deep hedging presents an enhanced alternative to traditional sensitivity-based methods in real markets, a particular challenge of this task is the relatively limited availability of financial data (e.g., stock prices) since an extensive amount of input data is usually required to calibrate neural networks. Moreover, even if a large amount of historical data is available, predicting future trends, future variance, or the future distribution remains a highly complex task.
\\
Various studies have considered synthetic training paths, which aim at resembling the actual financial market dynamics and are generated by a single stochastic model with fixed parameters to investigate the performance and behaviour of the Deep Hedging framework. For instance, \cite{Buehler2018} uses data generated by the Heston stochastic volatility model, while \cite{horvath2021deep} applies the rough Bergomi model. However, in financial markets the underlying dynamics change quite frequently, leading to a relative high parameter uncertainty and model risk. An approach towards the inclusion of this uncertainty in a deep learning approach for hedging is offered by Lüthkebohmert et al. (2021). The authors suggest a deep hedging approach under parameter uncertainty for generalized affine processes and show their neural network model's robustness against changes in the dynamics of the underlying. Nevertheless, the performance of neural network models outside their training regime can be quite poor, such that in order to learn a new task (i.e., a hedging strategy under paths not seen in training process) it would be desireable to train a new neural network from scratch. This can become relatively quick computationally expensive. Moreover, given that models in this context typically undergo extensive validation, requiring significant human effort, the process of recalibrating or training a new model from scratch can be resource-intensive and costly in this regard beneath the computational aspects.
\\
Recently, it has been recognized that multi-task learning and especially task embedded networks can remedy this situation. In particular, it has been demonstrated, that this approach performs well in the context of solving problems in finance, such as learning a family of models for given data an then recalibrate the respective parameters to new data, and it has also been successfully applied in the context of time series forecasting (e.g. \cite{schreiber2021task}), natural language processing (cf., \cite{chen2024multi}), and many other applications (see for instance \cite{caruana1996algorithms}). 
More specifically, task embedded networks can help to utilize a trained neural network outside of the parameter range where it was originally trained by encoding specific tasks using an embedding layer, which is then combined with other input features. Thus, task embedded networks make use of the previously learned neural network while adjusting it relatively fast and robust for a new parameter regime an therefore provide an effective approach with reduced computational effort (compared to other MTL architectures). 
\\
\noindent
With the motivation and the background described above the plan of the work reported here has been to apply a simple multi-task learning architecture to learn a hedging strategy under model uncertainty.
To this end, we investigate the performance of the Deep Hedging framework beyond using training paths generated by just one model. In particular, we implement and analyze a task embedded neural network, which is able to derive hedging strategies for market dynamics with different properties than that of the training dataset.
\\
The article is structured as follows: Section 2 provides an overview of the Deep Hedging framework under model uncertainty and describes the design of our neural network. Section 3 presents several numerical experiments that provide insights into the behaviour and performance of the framework. A conclusion and brief outline for future work is given in section 4.

%-----------------------------------------------------------
\vspace{-0.3cm}
\section{Methodology}

In this section, we provide a compact description of the deep hedging framework, and discuss our particular implementation of this approach for hedging under model uncertainty. Note that for mathematical details about the original deep hedging framework, the reader is referred to \cite{Buehler2018}. 

\vspace{-0.3cm}
\subsection{PnL modelling}

We assume a discrete time financial market with $d$ hedging instruments (e.g., stocks), which are denoted by the stochastic process $S$, a finite time horizon $T$ and $n$ discrete trading dates $0 = t_0 < t_1 < \ldots < t_n = T$.
\\
The objective is now to hedge against a given contingent claim $Z$, which is completely known at $T$, using a hedging strategy $\delta = (\delta_k)_{k=0,\ldots,n-1}$. In particular, $\delta_k\in\mathbb{R}^d$ describes the amount of $d$ hedging instruments in the portfolio $V$ (owned by the hedger) at time $t_k$.
The value of the portfolio for $Z$ at maturity $T$ is then given by
\begin{align}
\label{PnL}
\mathrm{PnL}_T(p_0,Z, \delta) = Z + p_0 + (\delta \cdot S)_T ,
\end{align}
%where 
\begin{align}
(\delta \cdot S)_T = \sum_{k=0}^{n-1} \delta_k \cdot (S_{k+1} - S_k)
\end{align}
and $p_0 > 0$ is the exogeneously given fair price (i.e., $p_0$ is calculated separately from our described method). 
\\
In the case of no transaction costs, we solve the following optimization problem to find an variance optimal hedging strategy for $Z$ using a vector of hedge instruments (i.e., we want to find the best hedging strategy $\delta$ to minimize the losses of the PnL of a given portfolio):
\begin{align}
\label{loss_mean_var}
\mathrm{inf}_{\delta \in \mathcal{H}} \mathbb{E} \left[ ( Z +  (\delta \cdot S)_T)^2 \right] ,
\end{align}
where $\mathcal{H}$ summarizes all valid hedging strategies. Note that instead using variance as a measure of hedge quality, we could also use utility functions or convex risk measures within the cost function, see \cite{Buehler2018}.
\\
To solve the optimization problem (\ref{loss_mean_var}), numerically in the framework of a neural network we replace as in \cite{Buehler2018} the set of all valid hedging strategies by a certain subset $\mathcal{H}_\Theta  \in \mathcal{H}$ whose hedging strategies $\delta_\Theta$ depend on a discrete number of neural network parameters $\Theta \in N, N >0$ (i.e., network weights as parametrization for hedging strategies), such that equation (\ref{loss_mean_var}) becomes
\begin{align}
\begin{split}
\mathrm{inf}_{\delta \in \mathcal{H}_\Theta} \mathbb{E} \left[ ( Z +  (\delta \cdot S)_T)^2 \right] \\ =  \mathrm{inf}_{\theta \in \Theta} \mathbb{E} \left[ ( Z +  (\delta \cdot S)_T)^2 \right] .
\end{split}
\end{align}
\\
In \cite{Buehler2018}, the utility of this hedging approach is shown for Black-Scholes (BS) and Heston models. However, in each case only a single model with fixed parameters is considered. To achieve a deep learning approach for hedging option portfolios under model uncertainty, we have extended this framework by a method that is based on key ideas from word embeddings in natural language processing, and is along the lines of the very simple neural network architecture for multi-task learning as introduced in \cite{schreiber2021task} and applied by \cite{oeltz2023parameterizedneuralnetworksfinance} to solve problems in finance (i.e., the calibration of spread curves). The latter highlighted in a suite of numerical experiments the utility of the task-embedded neural network to learn based on a set of different models for given data rather than just learning for a specific data sample generated by a single model, which makes it relatively easy to recalibrate model parameters to new data. We adjust this method to learn hedging strategies and include aspects from Lüthkebohmert et al. (2021) for pricing and hedging under uncertainty. In the following we give a very quick overview of the idea behind multi-task learning and discuss a particular, simple neural network architecture of this method using an embedding layer for deep hedging. 

\vspace{-0.3cm}
\subsection{Neural network with Embedding}

In multi-task learning, there are generally two concepts for sharing knowledge between tasks: soft parameter sharing and hard parameter sharing. Within the former approach, each task has its own neural network model and similarities between tasks are maintained through regularization techniques, while the latter approach involves using a shared set of layers for all tasks, with some specific layers dedicated to individual tasks (and therefore, requiring less parameters compared to soft parameter sharing). In this study, following \cite{oeltz2023parameterizedneuralnetworksfinance}, we focus on a relatively simple multi-task learning architecture, which belongs to the category of hard parameter sharing and is called task embedding network. Note, that  we refer to this in the following of this article as \textit{parameterized neural network} (PNN). 
\\
The general idea of multi-task learning can be formulated using $m$ learning tasks $\mathcal{T}_{i=1}^m$, which are accompanied by a training dataset consisting of $n_i$ training samples, respectively. Following \cite{zhang2021survey}, a definition of multi-task learning then reads: \textit{Given $m$ learning tasks} $\mathcal{T}_{i=1}^m$, \textit{where all the tasks or a subset of them are related, multi-task learning aims to learn the $m$ tasks together to improve the learning of a model for each task} $\mathcal{T}_i$  \textit{by using the knowledge contained in all or some of other tasks.}
\\
To further illustrate this approach and especially the PNN architecture, we schematically visualize its concept in terms of learning a hedging strategy in Figure 1 and describe how this approach can help to utilize a neural network outside the parameter regime (i.e., within a family of parameterized models) where it was originally trained on, which can be a particularly challenging task in the context of stock price dynamics. 
%\vspace{-0.3cm}
\begin{center}

%\begin{wrapfigure}{l}{\linewidth}
%\begin{figure}
%
\includegraphics[scale=.39]{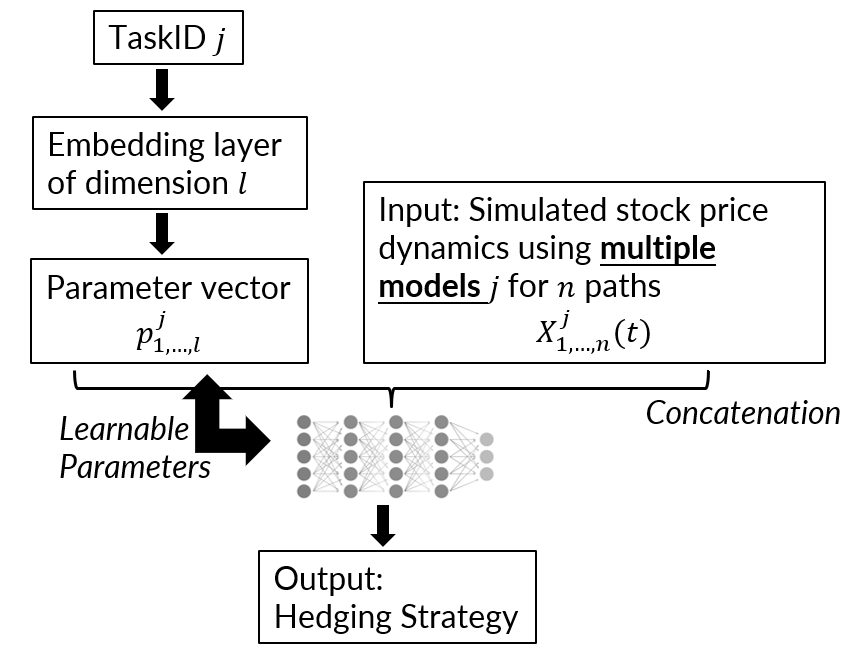}
{ Figure 1: Parameterized network architecture.}
\label{fig:PNN}
%\end{wrapfigure}   
\end{center}
%\noindent 

Suppose, that we have as an input a training dataset $\mathcal{D}_{source}$, which consists of simulated stock price dynamics from a class of $m$ different models, i.e., with each model $1 \leq j \leq m$ we simulate $n$ sample paths for the time dependent state variable. Note, that for simplicity we consider the same number of sample paths for each model. Then, a task, such as a hedging strategy consistent with $\mathcal{D}_{source}$, can be learned using a neural network. However, there may be a need to learn a new task (like solving a hedging problem) within $\mathcal{D}_{target}$, which does not align with the model parameters or assumptions in $\mathcal{D}_{source}$ and which consists of data samples from a different model, such that $\mathcal{D}_{target} \neq \mathcal{D}_{source}$.
\\
Deep learning with PNNs aims to enhance the new task by leveraging knowledge from $\mathcal{D}_{source}$. Without embedding, the neural network trained on $\mathcal{D}_{source}$ would be discarded, requiring a new neural network to be trained from scratch with a large amount of data from $\mathcal{D}_{target}$. PNNs avoid this by utilizing information from the $\mathcal{D}_{source}$ and using only a small amount of data from $\mathcal{D}_{target}$ to modify the parameters of the previously trained neural network. 
\\
Thereby, the $m$ different models are encoded by (discrete) \textit{TaskIDs} (i.e., integers $j$), assigned as input to the embedding layer. Then an embedding layer of dimension $l$ is utilized to transform these integers into a corresponding model-specific parameter vector $p_i \in \mathbb{R}^l$ for each model. By concatenating the results of the embedding space with the other input data, the input for the main neural network, which remains consistent across all models in terms of the network weights, is defined. During training, each parameter vector is fine-tuned by optimizing the weights of the embedding layer. 
\\
If we assume that we have to calibrate the neural network to additional training data (i.e., new data from a new model is available, which is encoded by a new TaskID), then one simply needs to find a new parameter $p_{new} \in \mathbb{R}^l$ and get a model for this task. Note that, for simplicity, we chose the start value for the parameter as the average over all parameters that have been calibrated so far.
As only the low-dimensional parameter vector is adjusted, the risk of overfitting is significantly mitigated. Therefore, this approach remains applicable even when only a limited training dataset is available, making it particularly advantageous in our case, as it enables adaptation to changing market conditions through recalibration using only the most recent data. Furthermore, if the overall model, including the parameter vector, has already undergone comprehensive validation and testing, an extensive re-evaluation after recalibration may not be necessary. This can significantly reduce both the time-to-market and the associated human effort.
\vspace{-0.3cm}
\section{Models}

In this article, simulations of time dependent stock price processes $S = \{ S(t) , 0 \leq t \leq T \}$ are performed by four different models, i.e., the geometric Brownian motion, the Heston Stochastic Volatility model and its generalization allowing for jumps as well as the Barndorff-Nielsen-Shephard model (cf. \cite{schoutens2003perfect}). In the following we briefly describe their stochastic differential equations.

\vspace{-0.3cm}
\subsection{Geometric Brownian Motion}

The geometric Brownian motion (GBM) is a widely-used continuous-time stochastic model to describe stock prices in the BS model. 
Its stochastic differential equation takes the following form
\begin{align}
\mathrm{d} S = \mu S \mathrm{d} t + \sigma S \mathrm{d} W ,
\end{align}
where $W$ is a one-dimensional Brownian motion. The parameters $\mu$ and $\sigma$ denote the drift and volatility constants.

\vspace{-0.3cm}
\subsection{Heston Stochastic Volatility Model}

Next, we consider stochastic volatility models, where the volatility is allowed to be a random variable themselves and is additionally governed by a stochastic differential equation, such that the simulation of the stock price process is one step closer to reality.
The Heston model is the most classic stochastic volatility model and its system of stochastic differential equations is given by
\begin{align}
\mathrm{d} S =  \sigma S \mathrm{d} W ,
\end{align}
\begin{align}
\mathrm{d} \sigma^2 = \kappa (\eta - \sigma^2) \mathrm{d} t + \theta \sigma \mathrm{d} \tilde{W} ,
\end{align}
where $W$ and $\tilde{W}$ are two correlated one-dimensional Brownian motions with $\mathrm{Cov} [\mathrm{d} W \mathrm{d} \tilde{W}] = \rho \mathrm{d} t$. Note, that the variable $\sigma$ now controls the volatility of $X$ by a mean-reverting stochastic process instead of a constant. Thereby, the parameters $\kappa, \eta$ and $\theta$ denote the rate of mean reversion, the long-run average variance of the price and the volatility of the volatility (i.e., the variance of $\sigma$), respectively.

\vspace{-0.3cm}
\subsection{Heston Stochastic Volatility Model with Jumps}

We also consider an extension of the Heston model by a jump process in the stock prices, where jumps occur as a Poisson process. Hence, the stochastic differential equation for $S$ is supplemented by a Poisson process  $N = \{ N(t) , 0 \leq t \leq T \}$, which is independent of $W$ and $\tilde{W}$ and has an intensity parameter $\lambda > 0$, such that $\mathbb{E}(N) = \lambda t$:  
\begin{align}
\mathrm{d} S = \ldots - \lambda S \mu_J \mathrm{d} t +  S J \mathrm{d} N ,
\end{align}
while the stochastic differential equation for the volatility remains unchanged. Here, $J$ denotes the jump size, which is lognormally, identically and independently distributed over time with unconditional mean $\mu_J$ and standard deviation of $\mathrm{log} (1 + J)$ as
\begin{align}
\sigma_J = \mathcal{N} \left( \mathrm{log} (1 + \mu_J) - \frac{\sigma_J^2}{2} , \sigma_J^2  \right) ,
\end{align}
where $\mathcal{N}$ denotes a Normal distribution.

\vspace{-0.3cm}
\subsection{Barndorff-Nielson-Shephard Model}

Finally, the Barndorff-Nielson-Shephard (BNS) model is specified by the stochastic differential equations
\begin{align}
\mathrm{d} Z =  (- \lambda k(-\rho) -  \frac{\sigma^2}{2}) \mathrm{d} t + \sigma  \mathrm{d} W + \rho \mathrm{d} z_{\lambda t} ,
\end{align}
\begin{align}
\mathrm{d} \sigma^2 =  - \lambda \sigma^2 \mathrm{d} t + \mathrm{d} z_{\lambda t} ,
\end{align}
where $\lambda$ describes a positive constant, $Z = \mathrm{log} (S)$ is the log-price process and the volatility is modeled by an Ornstein Uhlenbeck process driven by a compound Poisson process, i.e., 
\begin{align}
z_t = \sum_{n=1}^{N} x_n ,
\end{align}
with a Poisson process  $N = \{ N(t) , 0 \leq t \leq T \}$ which has an intensity parameter $a$ such that $\mathbb{E}(N) = a t$ and $\{ x_n, n = 1,2 \ldots \}$ is an independent and identicall distributed sequence, where each $x_n$ follows an exponential law with mean $b^{-1}$. Moreover, the constant $\rho$ describes the correlation between the log-price and the volatility and the function $k(u)$ denotes the cumulant generating function of $z_1$ and is defined by
\begin{align}
k(u) = \mathrm{log} \mathbb{E} [ \mathrm{exp} (-u z_1)] = \frac{-au}{b+u} .
\end{align}

\vspace{-0.3cm}
\section{Numerical experiments}\label{sec_results}
% Alle Auswertungen in diesem Abschnitt wurden mit dem Skript create_figures.py erstellt
In this section, we study the accuracy and efficiency of our proposed deep hedging framework under model uncertainty. If not stated otherwise, we specify the setting of the PNN as follows:
The neural network has 3 inner layers with 128 neurons each and we apply to all inner layers $SELU$-activation functions. The algorithm is coded in Python using the Tensorflow-environment, in which we execute the algorithm with a batch-size of 1024, an Adam optimizer with 1000 epochs and an exponentially decaying learning rate with an initial value of $0.0005$.
The learning rate is reduced during training by an exponential schedule so that the final learning rate equals $0.0001$.

\vspace{-0.3cm}
\subsection{Results under a family of GBM models}
In our first case study, the goal is to hedge a short at-the-money European call option position with strike $K = 1$, $S_0 = 1$ and payoff $(S_T - K)^+$ for $T = 30/365$, daily rebalancing (i.e., $n = 30$) and trading dates $t_i = i/365$ for $i = 0, \ldots, n$. For simplicity, only $d = 1$ hedge instrument is considered in the market (i.e., the underlying stock price $S_t$) and we simulate $S_t$ under the GBM framework.
\\
Given our objective of developing a hedging strategy capable of adapting to new market regimes using limited training data, we refrain from relying on a single model with fixed parameters. Instead, we compose the training dataset using paths simulated for several GBM models with different volatility parameters, i.e., we set $\mu = 0$ and sample the volatility parameter such that $\sigma$ is uniformly distributed on $[0.1, 0.8]$. Because the considered stock price process is used to model stock prices in the BS framework, it offers a possibility to compare the results of our PNN to the analytically calculated solutions.
%Moreover, in this experiment we analyze the performance of the method in relation to the number of simulated paths per task. 
\\
First, we analyze the performance of our PNN in relation to the number of training tasks and the number of simulated paths per task used in the training data.
\vspace{-0.3cm}
\begin{center}
%\begin{figure}
\includegraphics[scale=.55]{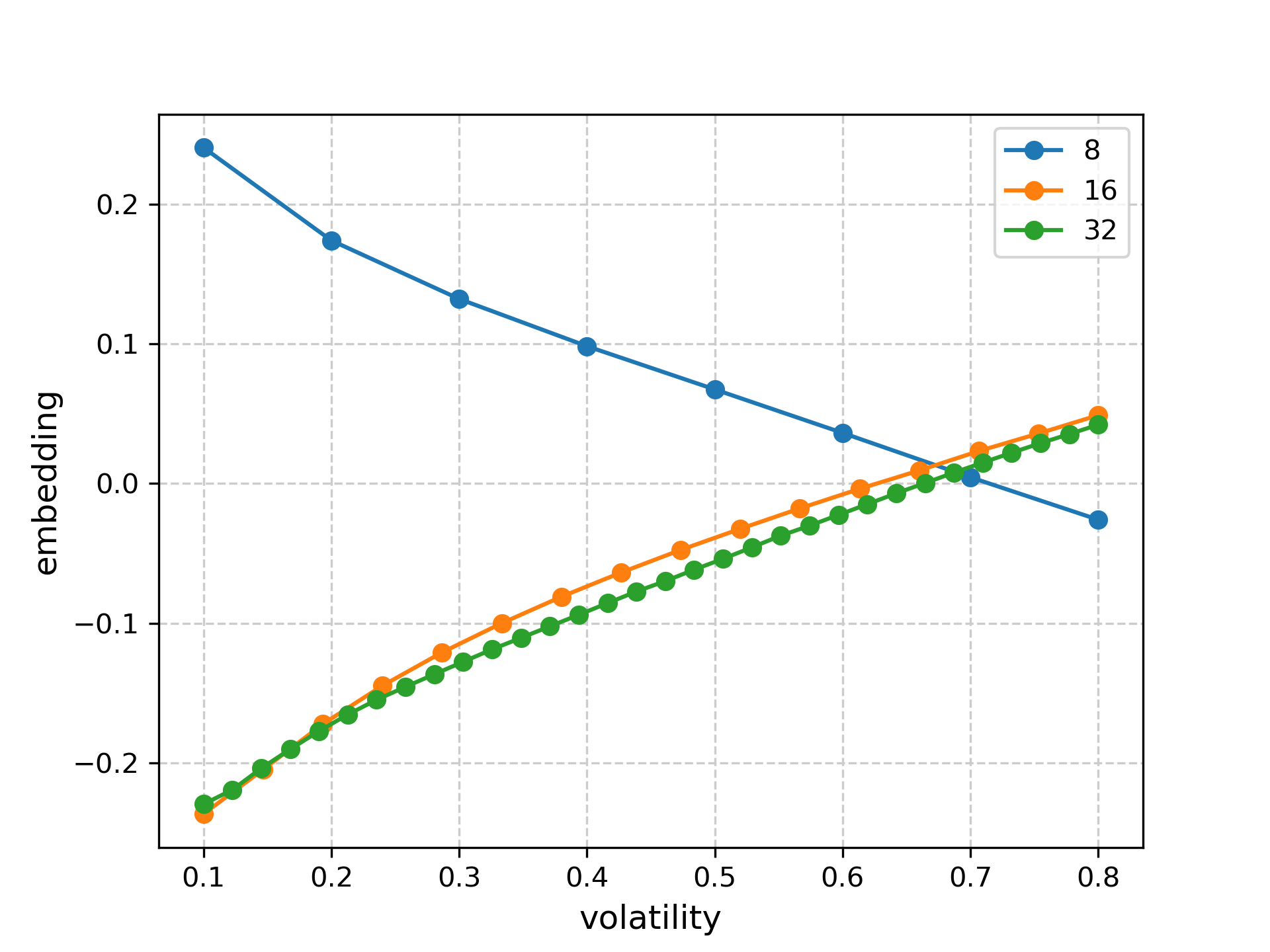}  \\
{ Figure 2: Volatility of single GBM tasks versus learned one-dimensional embedded value of the respective task.}
%\end{figure}   
\end{center}
\noindent 
In Figure 2 we depict for several tasks (i.e., the training data consists of 8 (blue), 16 (orange) and 32 (green) different GBM models with $\mu = 0$ and volatility parameters $\sigma \in [0.1, 0.8]$) the corresponding parameter vector as a scatter plot. We clearly see a (almost linear) monotonic dependency between the volatility parameter and the embedding. Moreover, we observe that an increasing number of different tasks does not affect the structure of the resulting embedding significantly.
\vspace{-0.5cm}
\begin{center}
%\begin{figure}
\includegraphics[scale=.55]{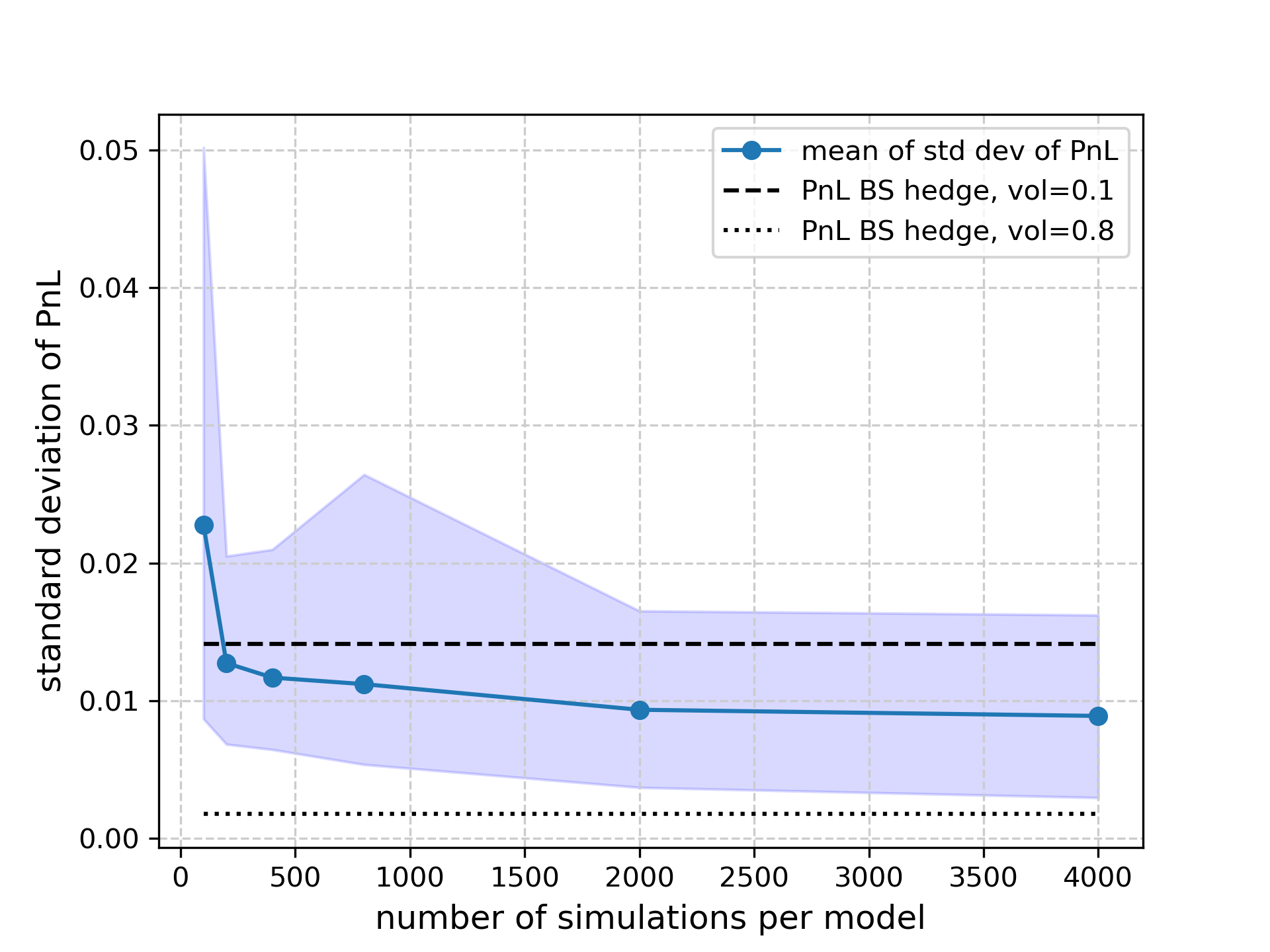} \\
{Figure 3: Range and mean of the standard deviation over all tasks versus the  number of simulations per task.}
%\end{figure}   
\end{center}
\noindent 
Figure 3 shows the range of the standard deviation of the PnL distributions (including the mean of the standard deviations) over all tasks for an increasing number of simulation paths per task. As a baseline, we also plot the standard deviation of the PnL for a BS hedge on GBM paths with $\sigma = 0.1$ and $\sigma = 0.8$. For a smaller number of simulation paths, the mean of the standard deviation is slightly larger than for a higher number of simulations per model. Moreover, for a high number of simulations (i.e., $> 2000$) per model the maximum and minimum standard deviation of the PnL over all tasks is nearly equal to the corresponding standard deviation of the PnL for a BS hedge. 
\vspace{-0.7cm}
\begin{center}
%\begin{figure}
\includegraphics[scale=.55]{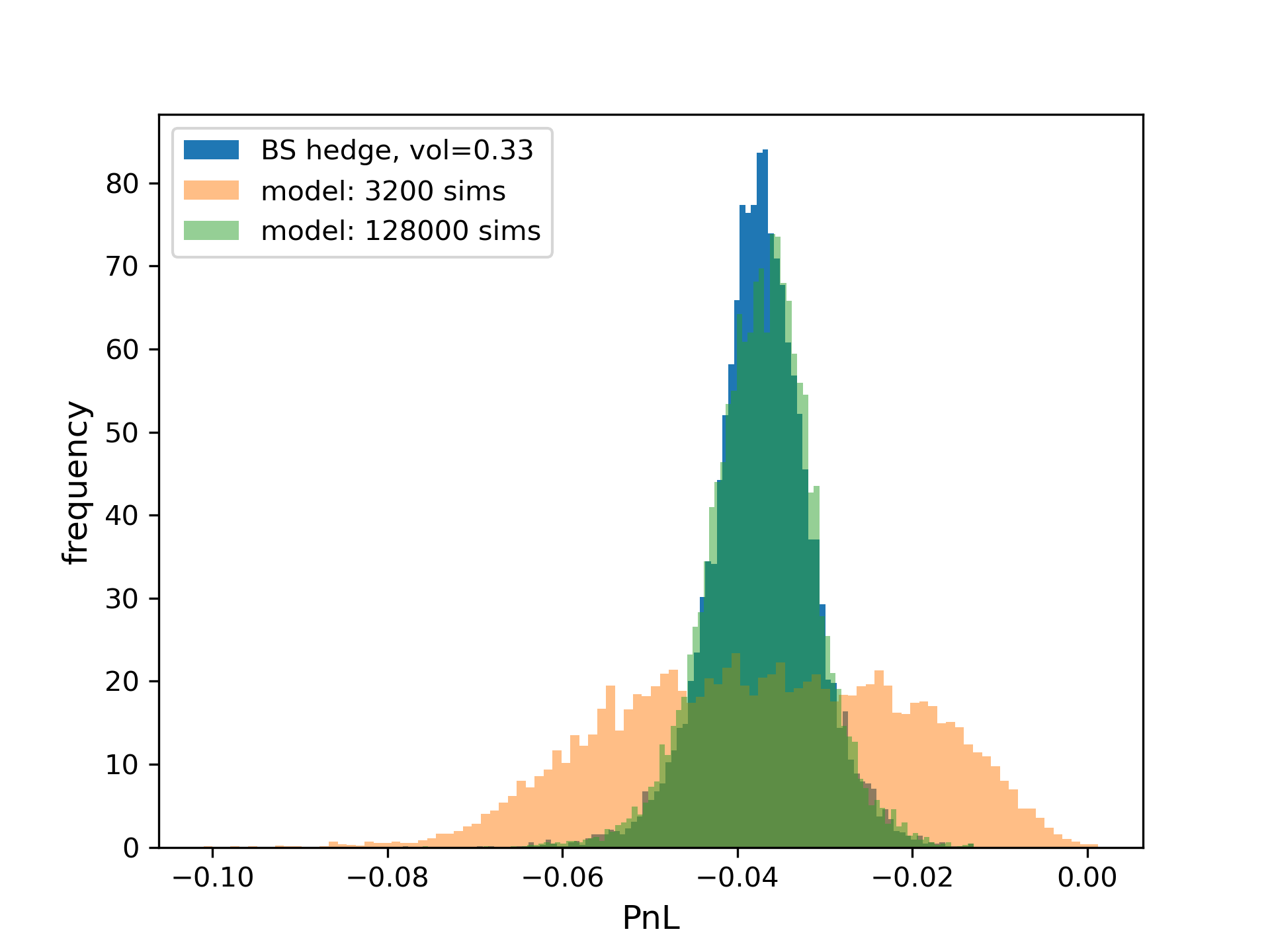} % \\
{Figure 4: PnL Distribution for a BS hedge on GBM paths with $\sigma=0.33$ in comparison to hedge performance of multi-task deep hedging models trained on  100 and 4000 simulated paths per training task.}
%\end{figure}   
\end{center}
\vspace{-0.1cm}
\begin{center}
%\begin{figure}
\includegraphics[scale=.55]{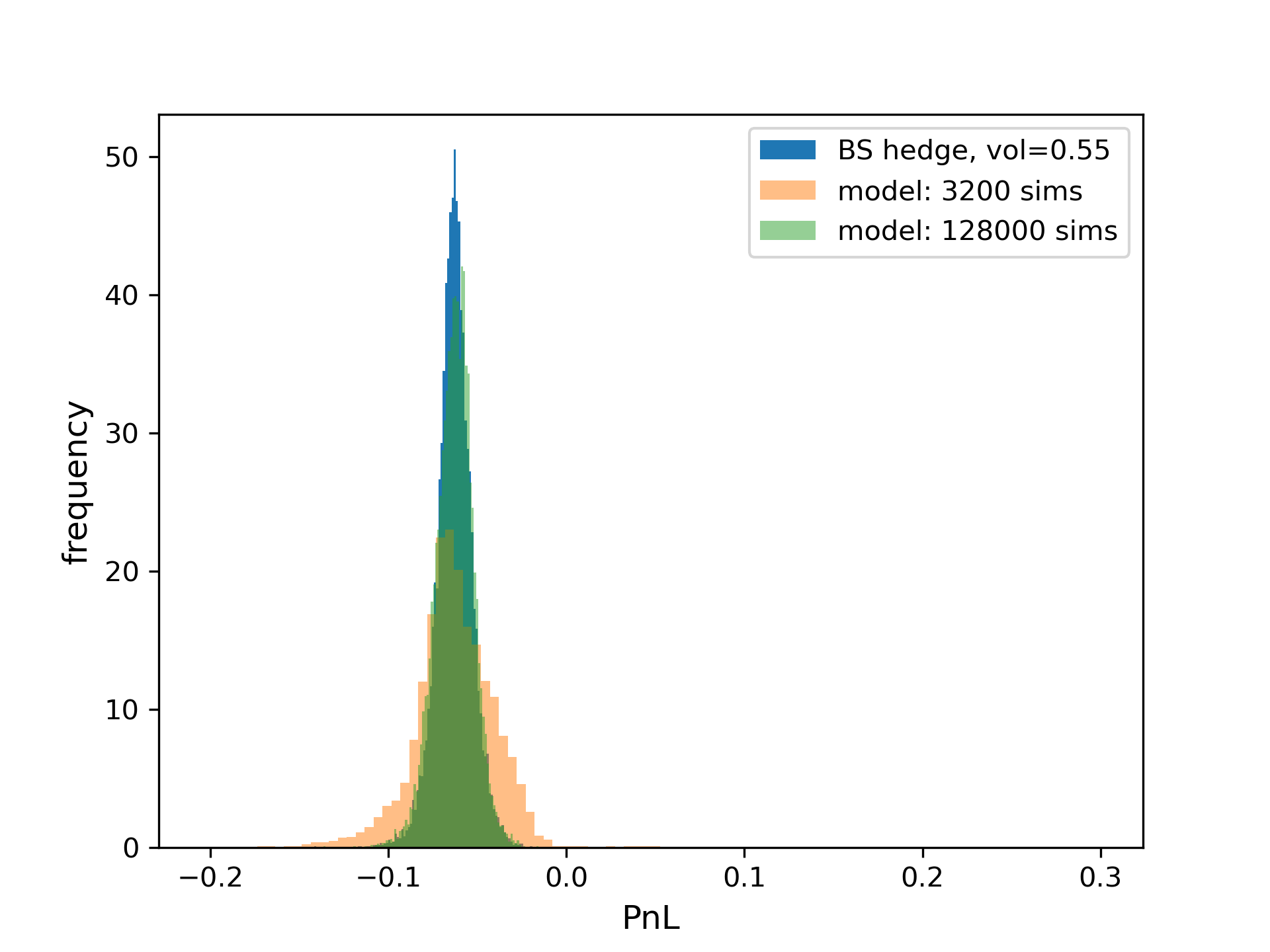}
{Figure 5: PnL Distribution for a BS hedge on GBM paths with $\sigma=0.55$ in comparison to hedge performance of multi-task deep hedging models trained on  100 and 4000 simulated paths per training task.}
%\end{figure}   
\end{center}
\noindent 
Figures 4 and 5 show some examples of the resulting PnL histograms from our PNN trained on 32 tasks compared to the BS strategy for two selected volatility parameters, $\sigma=0.33$ and $\sigma=0.55$ (i.e., two selected tasks). We observe, that our PNN result trained with a large number of simulations paths is much closer to the BS strategy and produces similar results in terms of the PnL distribution than the neural network model trained on a relative small number of simulated paths.
\end{multicols} % End the three-column layout for a large picture
%\newpage

\begin{figure}
\renewcommand{\thefigure}{6}
    \centering
    \includegraphics[width=0.49\textwidth]{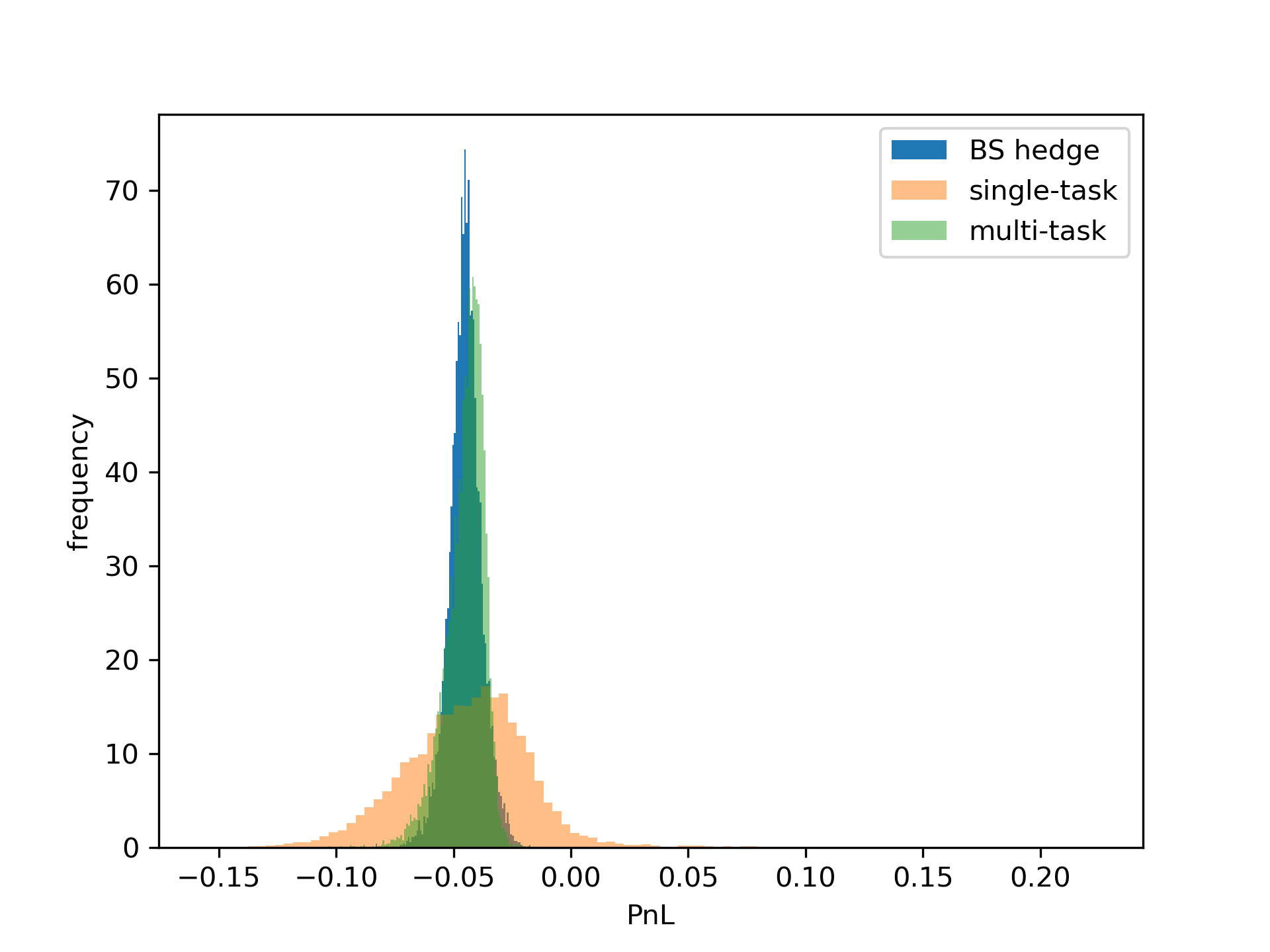 }
    \includegraphics[width=0.49\textwidth]{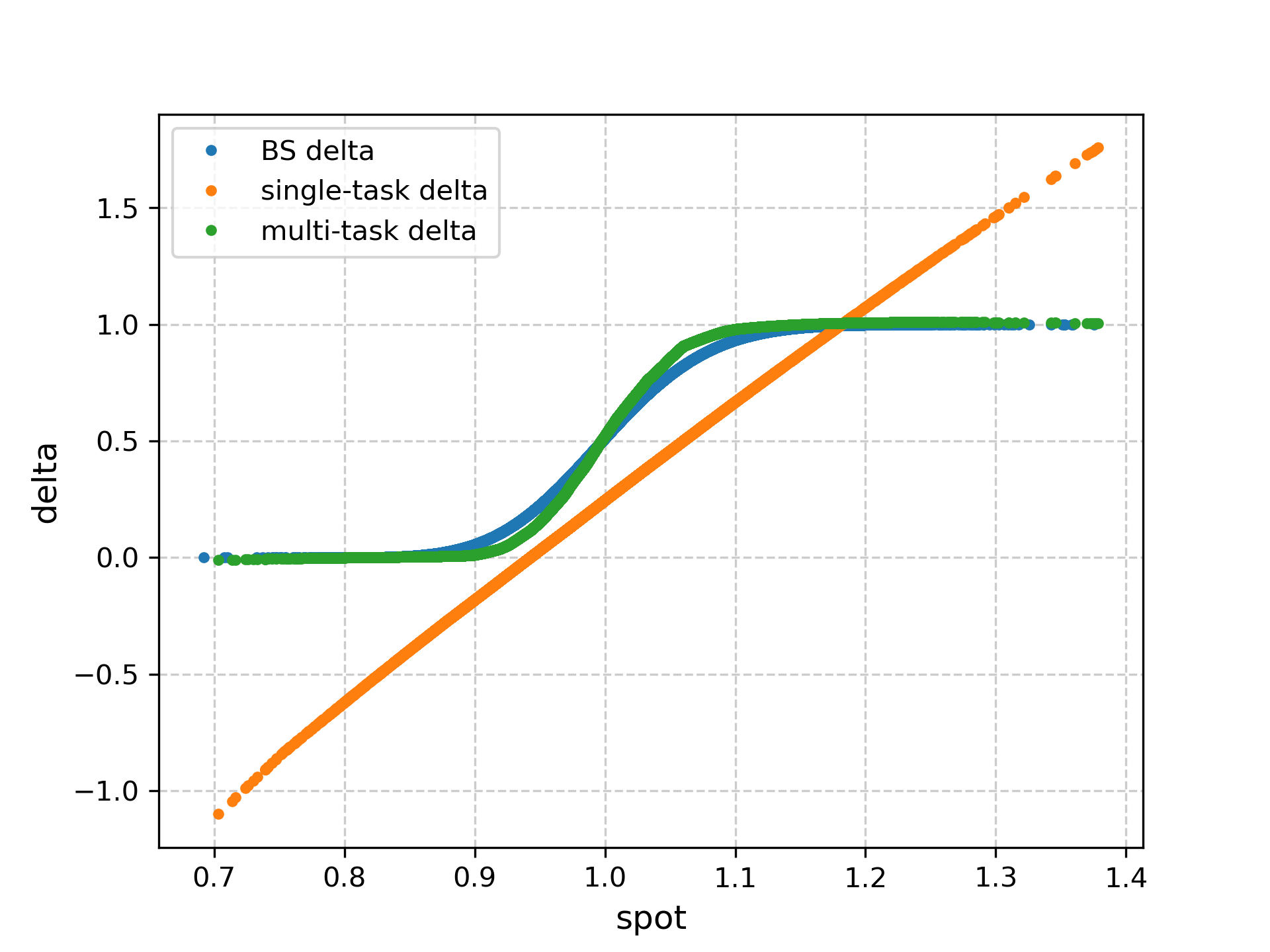 }
    \includegraphics[width=0.49\textwidth]{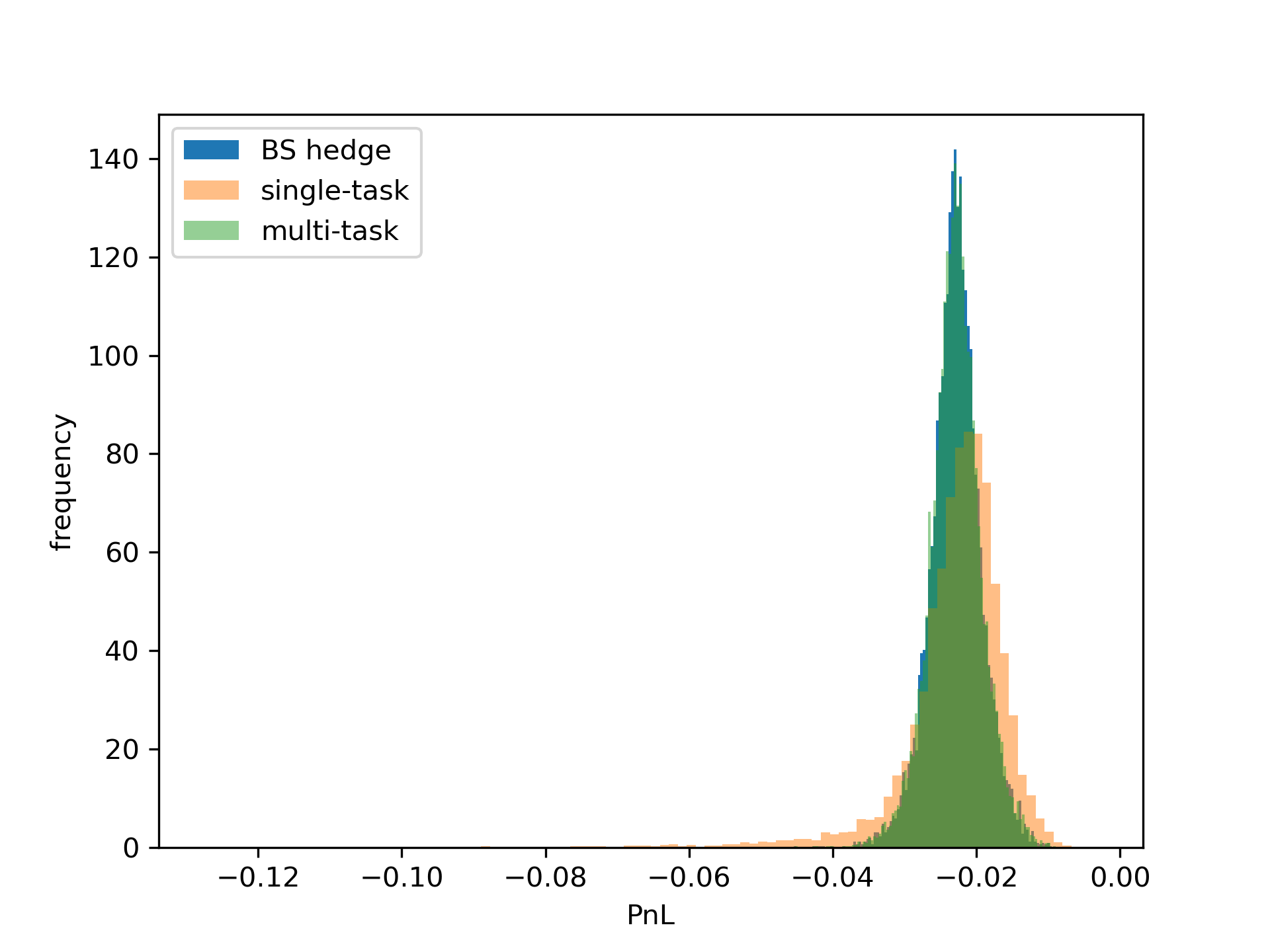 }
    \includegraphics[width=0.49\textwidth]{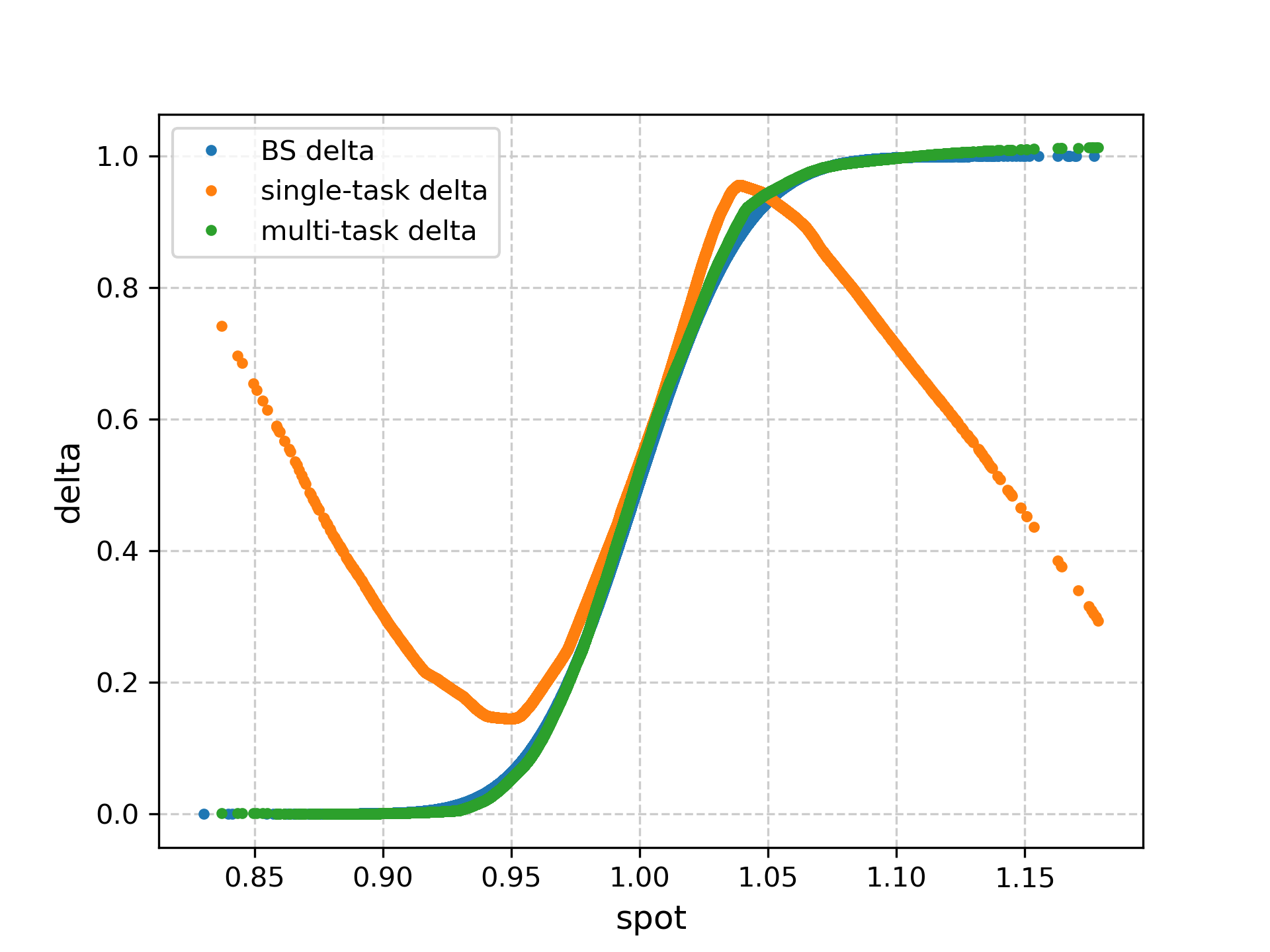 }
    \includegraphics[width=0.49\textwidth]{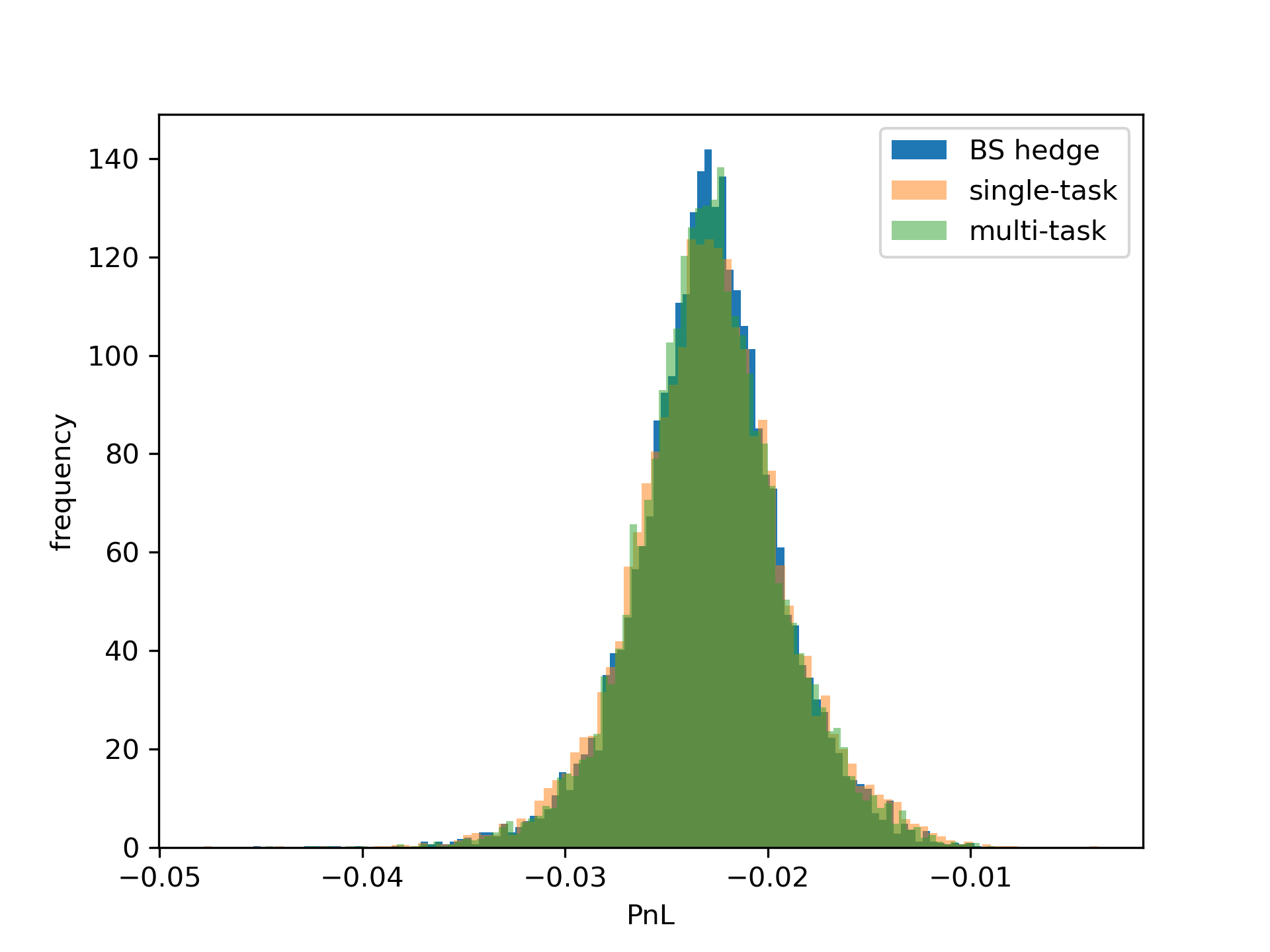 }
    \includegraphics[width=0.49\textwidth]{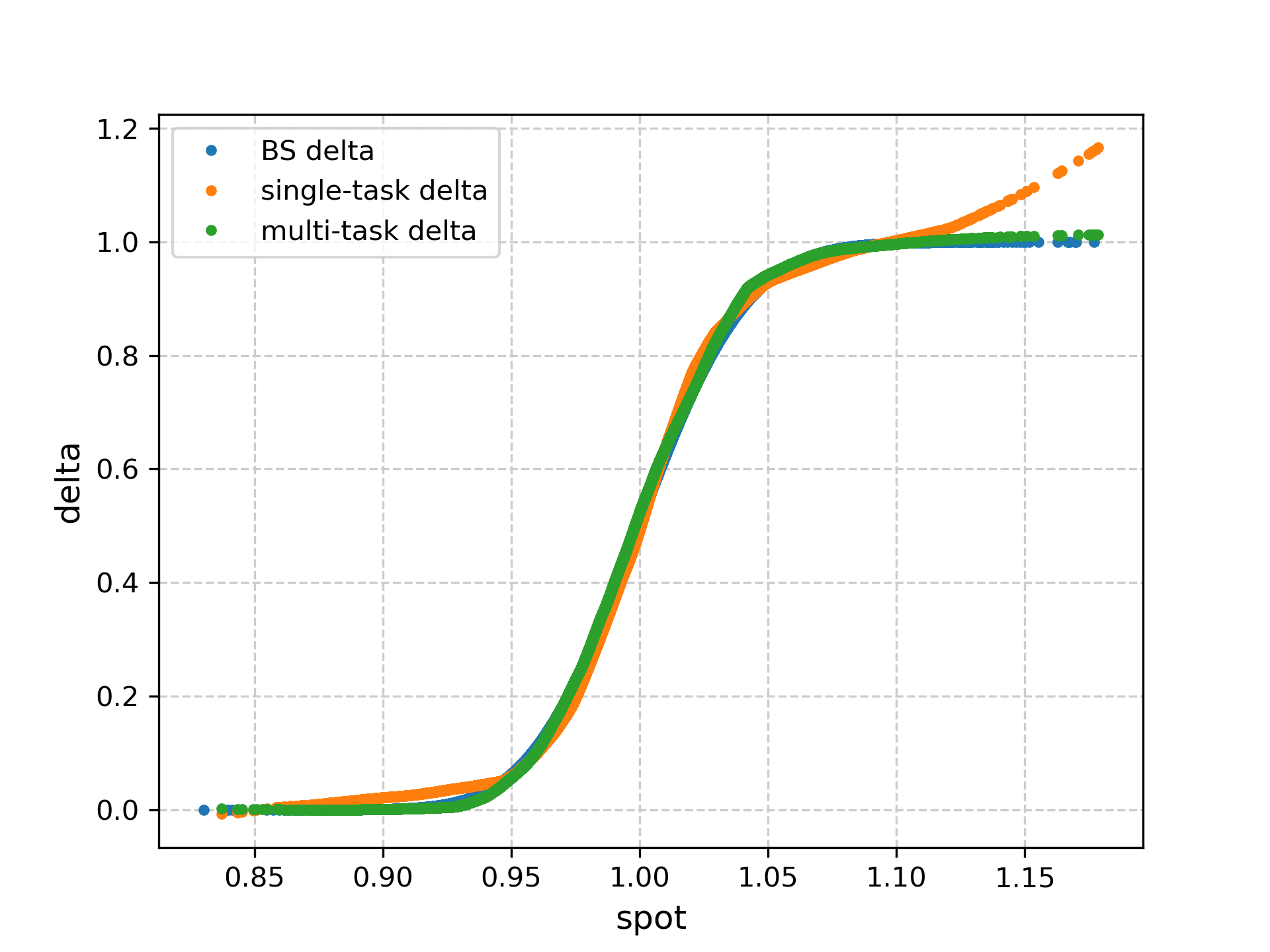 }
    \caption{ The figures on the left side compare PnL distributions for a BS Hedge, a single task network trained on 10 (top), 100 (middle) and 1000 (bottom) paths and the respective multi-task model where only the embedding vector is learned on the paths. The figures on the right side show the respective deltas ten days before option expiry.}
    \label{fig:dist_single_vs_multi}
\end{figure}

\begin{multicols}{2}

%\newpage
\end{multicols} % End the three-column layout for a large picture
\begin{center}
{Table 1: }{ Statistics of the PnL for parameterized deep hedging models trained on a different number of simulations. The training data consists of 32 different GBM models with uniform volatilities  between 0.1 and 0.8.}
\label{tab:table1}
\begin{tabular}{|c | c | c | c | c | c | c |} 
  \hline
 simulations/task &      mean &       std &   std min &   std max &  1\% quantile &  10\% quantile \\
  \hline
100    & -0.051158 &  0.035632 &  0.008580 &  0.050819 &     -0.150221 &      -0.091520 \\
200    & -0.051412 &  0.027267 &  0.006760 &  0.019854 &     -0.119152 &      -0.087161 \\
400   & -0.051381 &  0.026858 &  0.006385 &  0.020362 &     -0.118510 &      -0.085367 \\
800   & -0.051475 &  0.027449 &  0.005303 &  0.027408 &     -0.115599 &      -0.085771 \\
2000   & -0.051425 &  0.025895 &  0.003683 &  0.016086 &     -0.111582 &      -0.086758 \\
6400 & -0.051455 &  0.025430 &  0.001995 &  0.014037 &     -0.108649 &      -0.085949 \\
\hline
\end{tabular}
\end{center}
\vspace{0.1cm}
\begin{multicols}{2}

\noindent
In addition, Table 1 summarizes statistics of the numerical results of the PnL histograms depending on the number of simulations. We clearly see that the number of simulations from the training data influences the overall performance of the PNN. Especially, the standard deviation of the PnL decreases up to a factor of $1.4$ with an increasing number of simulations for the PNN. From these results we see, that the PNN is able to learn the structure and properties of the training data with model uncertainty. 
\\
Next, in Figure 6 we compare for a selected task the resulting PNN approximation (i.e., the multi-task model, where only the embedding vector is trained on the simulation paths) to results from a neural network trained on the single task only as well as the analytical solution of the corresponding BS strategy, and analyze the performance of the method in relation to the number of simulation paths on which the network was trained on. In particular, the PnL histograms are shown in the left panel of Figure 6, while the right panel displays the deep hedging strategies and the BS strategies $\delta$ ten days before option expiry as a function of the underlying price $S_t$ (i.e., $t = 20/365$).
\\
\noindent
 In summary, these results highlight the significant impact of the number of simulation paths on the performance of the learned strategies. When trained on $1,000$ simulation paths, both the single-task and multi-task networks yield results that closely align with the Black-Scholes (BS) benchmark strategy, exhibiting comparable profit-and-loss (PnL) distributions and well-matching performance curves. However, when the number of simulation paths is reduced, the parameterized neural network (PNN) demonstrates a superior ability to capture the structure of the target problem, even when evaluated on previously unseen data. This suggests that recalibrating only the parameter vector on new data not only accelerates the calibration process relative to conventional neural networks, but also leads to improved performance. These findings indicate that training the PNN on a diverse set of tasks—including simulated paths from various models and historical paths across different time periods and assets—yields a model family that can be adapted to new market regimes both efficiently and robustly.
 
\subsection{Results under a family of models with stochastic volatility}
\vspace{-0.3cm}
In this section, we evaluate our deep hedging framework with numerical experiments in a more complex framework, where we consider a family of models with stochastic volatility. Note that this is to be understood as a mere test of concept while further application and calibration to real-world data and a detailed analysis of the dependency of the observed PnL distributions on the parameter space is left to future studies. 
\\
%\vspace{-0.5cm}
\begin{center}
%\begin{figure}
%\vfill
%\begin{wrapfigure}{l}{\linewidth}
%scale=.55
\includegraphics[width=\linewidth]{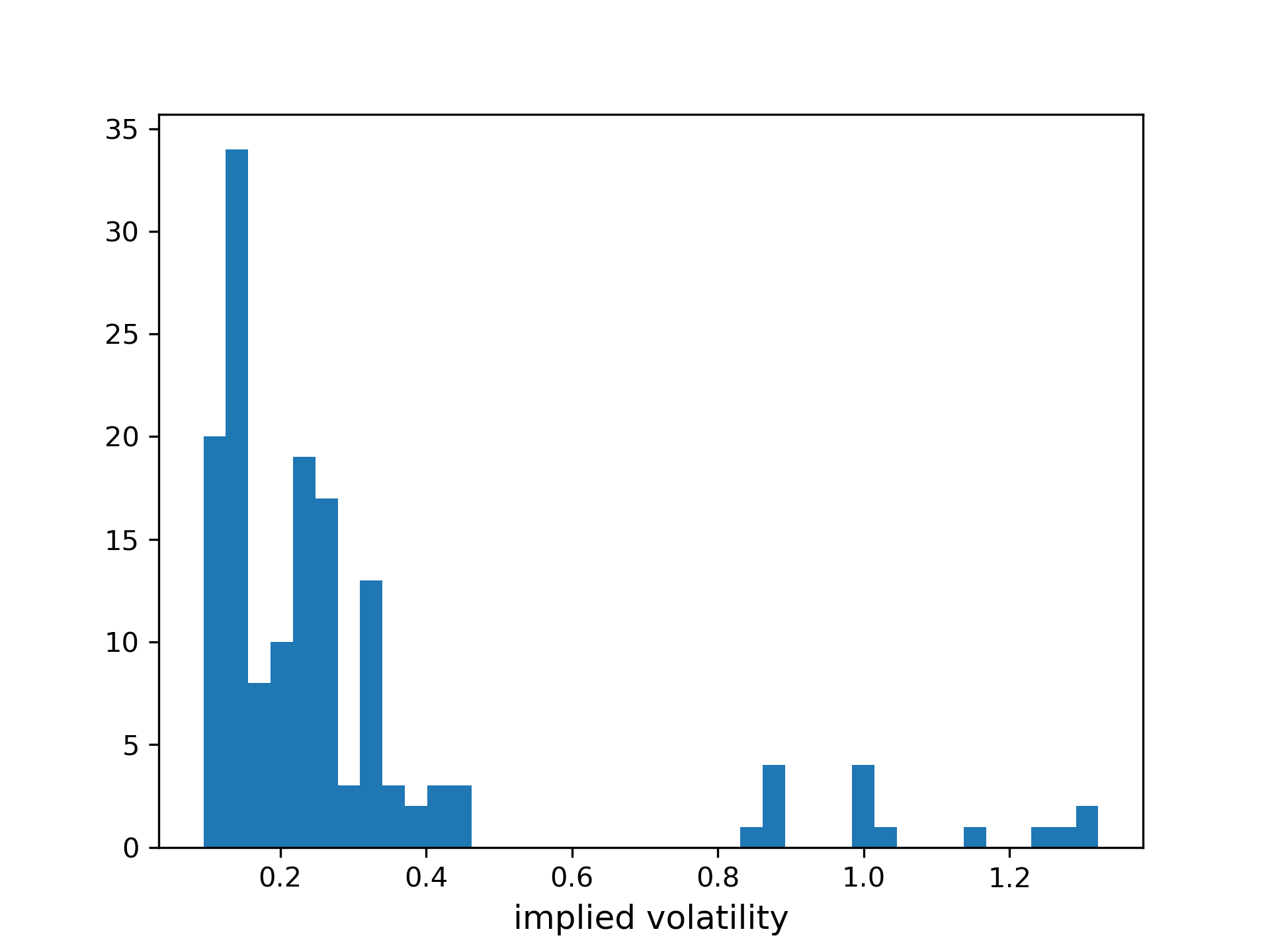 }
{Figure 7: Distribution of implied volatilities of the respective Call option evaluated by each of the stochastic volatility models.\label{fig:SV_Implied_Dist}}
%\end{wrapfigure}
%\end{figure}   
\end{center}
To this end, we consider model paths of the Heston stochastic volatility model, the Heston model with jumps and the Barndorff-Nielson-Shephard model, where each model has been calibrated to market option quotes of 50 different underlyings. In particular, we used option quotes for typical indices (such as DAX, STOXX50E, S\&P 500-index, CAC etc.) as well as option quotes on single equity indices. 
\vspace{-0.9cm}
\begin{center}
%\begin{figure}
\includegraphics[scale=.55]{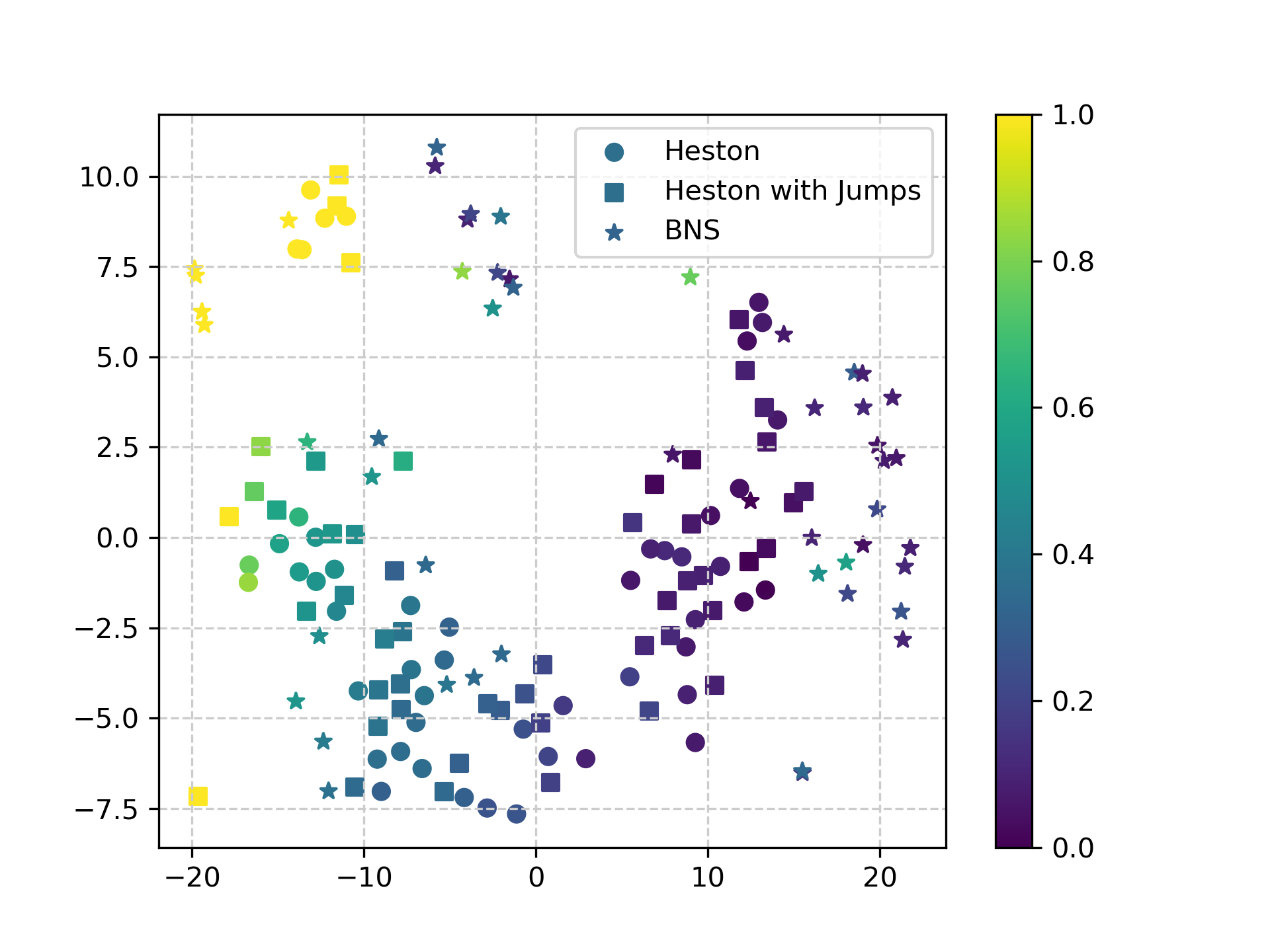 }
{Figure 8: t-SNE projected embedding vector from 8-dimensional embedding. The marker type encodes the respective model type, colour represents the ATM call price of the respective model parameters.}
%\end{figure}   
\end{center}
To give an impression on the range of values, we show in Figure 7 the implied volatility distribution of a European call option that expires in thirty days with ATM strike. 
Figure 8 illustrates, for several tasks, the high-dimensional embeddings produced by the parameterized neural network (PNN), visualized using the T-distributed Stochastic Neighbor Embedding (T-SNE) technique \cite{van2008visualizing}. The resulting structure reveals a clear organization of the learned parameter representations, which appear to correlate with the ATM call price associated with the respective model parameters, rather than being randomly or uniformly distributed. Additionally, Table 2 summarizes the statistical properties of the profit-and-loss (PnL) histograms across tasks. As benchmark strategies, we include the classical Black-Scholes (BS) delta-hedging approach, using the realized volatility computed across all simulated paths for each task, along with variants where this volatility is shifted by ±5 volatility points. The results indicate that the hedging strategy derived from our PNN framework consistently outperforms the BS benchmark, with the exception of the median PnL variance, which remains comparable between the two approaches. This demonstrates the effectiveness of the proposed PNN-based method under the specific scenario considered.

\end{multicols} % End the three-column layout for a large picture

\begin{center}
{Table 2: }{Statistics of variance of PnL for hedging on 40,000 simulated paths for each stochastic volatility model and three BS models with implied volatilities equal to the realized volatility over all paths as benchmarks. The Deep Hedging model's embedding dimension is eight and for the BS model delta we used the realized volatility (rlzd vol) over the respective paths and shifted volatilities where shift size is +5 and -5 volpoints (vp).}
\label{tab:table1}
\begin{tabular}{|c | c | c | c |} 
 \hline
 model &  mean variance &  median variance &  max variance \\
  \hline
  BS, rlzd vol - 5 vp&       3.98e-04 &         4.25e-05 &      3.98e-02 \\
    BS, rlzd vol&       3.87e-04 &         3.46e-05 &      3.92e-02 \\
  BS, rlzd vol + 5 vp&       3.87e-04 &         3.74e-05 &      3.87e-02 \\
       Deep Hedging &       2.94e-04 &         3.61e-05 &      2.52e-02 \\
        \hline
\end{tabular}
\end{center}

\begin{multicols}{2}

\vspace{-0.3cm}
\section{Summary}

The outcome of this study is a novel modeling framework for deep hedging that facilitates accelerated training in the presence of new market regimes while simultaneously enhancing robustness. This modeling approach is able to learn a family of models and portfolios, and allows for efficient recalibration of the respective parameters to new data avoiding overfitting. Closely related to the work of \cite{oeltz2023parameterizedneuralnetworksfinance}, we have complemented the deep hedging framework, originally designed by \cite{Buehler2018} for hedging a portfolio of derivatives using neural networks, for the best of our knowledge for the first time via task embedding. To that end, the neural network model has been modified to allow for encoding specific tasks using an embedding layer. 
For the verification that the new modeling framework is indeed accurate and efficient we have conducted idealized test cases. In particular, with a test case for hedging a short at-the-money European call option position under a family of GBM models we have validated the stability and accuracy of the code. It has been shown that the results of our task embedded framework compare well with analytical results from the BS framework. 
For a test of concept, we have also done simulations using a family of models with stochastic volatility. To keep the setting as simple as possible and allow for a precise measurement of the performance of the method we use artificially created data. The test simulations are promising and show that the proposed method is able to learn hedging strategies from a set of advanced models and can be stable and robustly recalibrated to new data. Nevertheless, for future studies it might be advisable to evaluate the performance of the deep hedging strategy with embedding on real financial data and include market frictions such as transaction costs.

\vspace{-0.3cm}
% \section{References}
% 1. Buehler, H. et al. (2019), Deep Hedging, https://arxiv.org/pdf/1802.03042
% \\
% 2. Caruana, R. (1996), Algorithms and Applications for Multitask Learning, https://citeseerx.ist.psu.edu/docu-\\ment?repid=rep1&type=pdf&doi=3980c955f95092e527c-\\580f9cfe066a17f752c08
% \\
% 3. Chen et al. (2024), Multi-Task Learning in Natural Language Processing: An Overview, https://arxiv.org/pdf/2109.09138
% \\
% 4. Horvath, B. et al. (2021), Deep Hedging under Rough Volatility, https://arxiv.org/pdf/2102.01962 
% \\
% 5. Luetkebohmert, E. et al. (2021), Robust deep hedging, https://arxiv.org/pdf/2106.10024
% \\
% 6. Oeltz, D. et al. (2023), Parameterized Neural Networks for Finance, https://arxiv.org/pdf/2304.08883
% \\
% 7. Ozbayoglu et al. (2020), Deep Learning for Financial Applications : A Survey, https://arxiv.org/pdf/2002.05786
% \\
% 8. Schreiber et al. (2022), Task Embedding Temporal Convolution Networks for Transfer Learning Problems in Renewable Power Time-Series Forecast, https://arxiv.org/pdf/2204.13908 
% \\
% 9. Schoutens, W. et al. (2003), A Perfect Calibration! Now What?, https://perswww.kuleuven.be/$\sim$u0009713/\\ScSiTi03.pdf
% \\
% 10. Zhang and Yang (2021), A Survey on Multi-Task Learning, https://arxiv.org/pdf/1707.08114
% \\
% 11. van der Maaten et al. (2008), Visualizing High-Dimensional Data Using t-SNE, Journal of Machine Learning Research 9:2579-2605, 2008.
\nocite{*}
\bibliography{literature.bib}
\end{multicols}
\vspace{-0.3cm}

%----------------------------------------------------------------------------------------

\end{document}